\documentclass[12pt]{article}
\usepackage[dvips]{graphicx}
\usepackage{epsfig}
\usepackage{amsmath,amssymb,amsbsy}
\usepackage{amsfonts}
\usepackage{bm}
\begin{document}
\thispagestyle{empty}
\vskip -3.6cm
~~~~~~~~~~~~~~~~~~~~~~~~~~~~~~~~~~~~~~~~~~~~~~~~~~~~~~~~~~~~~~~~CERN-PH-TH/2012-033\
\vskip 0.1cm
\begin{center}
{\Large\bf{High-energy asymptotic behavior of the \\
\vskip 0.6cm
Bourrely-Soffer-Wu model for elastic\\
\vskip 0.6cm
 scattering}}
\vskip 0.5cm
\vskip1.4cm
{\bf Claude Bourrely}
\vskip 0.3cm
Aix-Marseille Universit\'e,\\
D\'epartement de Physique, Facult\'e des Sciences de Luminy,\\
13288 Marseille, Cedex 09, France\\
\vskip 0.5cm
{\bf John M. Myers}
\vskip 0.3cm
School of Engineering and Applied Sciences\\
Harvard University, Cambridge, MA 02138, USA\\
\vskip 0.5cm
{\bf Jacques Soffer}
\vskip 0.3cm
Physics Department, Temple University\\
Barton Hall, 1900 N, 13th Street\\
Philadelphia, PA 19122-6082, USA
\vskip 0.5cm
{\bf Tai Tsun Wu}
\vskip 0.3cm
Harvard University, Cambridge, MA 02138, USA and\\
Theoretical Physics Division, CERN, 1211 Geneva 23, Switzerland\\
\clearpage
{\bf Abstract}\end{center}
Some time ago, an accurate phenomenological approach, the BSW model, was developed for
proton-proton and antiproton-proton elastic scattering cross sections at
center-of-mass energies above 10 GeV.  This model has been used to give
successful theoretical predictions for these processes, at successive collider
energies.

The BSW model involves a combination of integrals that, while computable
numerically at fairly high energies, require some mathematical analysis to reveal 
the high-energy
asymptotic behavior. In this paper we present a high-energy asymptotic
representation of the scattering amplitude at moderate 
momentum transfer, for the leading order in an expansion parameter closely related to
the logarithm of the center-of-mass energy.

The fact that the expansion parameter goes as the logarithm of the energy means that 
the asymptotic behavior is accurate only for energies greatly beyond any
foreseeable experiment. However, we compare the asymptotic representation against the
numerically calculated model for energies in a less extreme region of energy.
The asymptotic representation is given by a simple formula which, in particular, 
exhibits the
oscillations of the differential cross section with momentum transfer. We also
compare the BSW asymptotic behavior with the Singh-Roy unitarity upper bound for the 
diffraction peak.

\vskip 0.5cm

\noindent {\it Key words} : elastic scattering, total cross section, 
differential cross section\\

\noindent PACS numbers : 11.80.Fv, 13.85.Dz, 13.85.Lg, 25.40.Cm
\newpage
\section{Introduction}
Forty years ago, it was found theoretically, on the basis of quantum field
theory, that, contrary to the general belief at the time, the total cross
sections for hadronic scattering increases monotonically without limit at high
energies \cite{chengWu}. In order to make predictions that could be verified by
later experiments, it was essential to develop an accurate phenomenology with
the following characteristics:\\
(i) - it agrees with the above theoretical asymptotic result at high energies,\\
(ii) - it describes the experimental data at energies available at that time.

Such a phenomenological model, termed the BSW model, was formulated by three of
us \cite{BSW79,BSW84,BSW03}.

Our model belongs to a class of models attempting to describe 
with different approaches in the eikonal formalism the high energy behavior 
of p p elastic scattering,
let us mention a non exhaustive list \cite{block11}-\cite{petrov03}.

In this paper, after reviewing the BSW phenomenological model for the elastic
scattering amplitude, we present our predictions for LHC energies and derive
an asymptotic representation for the scattering 
amplitude for extremely high center-of-mass energies.

In section \ref{sec1} we recall the basic features of the impact picture 
approach (BSW), 
with its phenomenological parameters,
and the expression of the scattering amplitude used in the next sections.
In section \ref{sec2} we present our predictions for the LHC energies at
$\sqrt{s} = 7,~14\mbox{GeV}$ and make a comparison with the TOTEM preliminary
results.
The opacity function involves the evaluation of the function $F(x_\perp)$, 
we discuss
in section \ref{sec3} a detailed decomposition of its expression in view of
an asymptotic representation.
 In section \ref{sec4} we obtain the 
high-energy asymptotic representation of the BSW amplitude
in terms of a simple expression. This derivation involves a number of mathematical 
steps and the consideration of two different
kinematic regions. Some technical details are collected in the Appendix. 
Numerical calculations are presented in section \ref{sec5}, where we discussed some 
features of the real and imaginary parts of the scattering amplitude, generating some 
oscillations in the differential cross section. The comparison of the asymptotic 
representation 
and the exact BSW result is done at the LHC energy and also at a much higher 
center-of-mass 
energy of 6000TeV. In section \ref{conclu}, we also compare the asymptotic 
representation 
with the Singh-Roy unitarity upper bound for the diffration peak and we give our 
concluding remarks.

\section{The BSW model}
\label{sec1}
To describe the experimental data taken at the relatively low energies available
to experiments forty years ago, the BSW model was proposed, including Regge backgrounds. Both
for the energies of the present-day colliders and for the purpose of studying
the asymptotic behavior of the model at high energies, all the Regge backgrounds
can be neglected. The BSW model is given by the
following matrix element for elastic scattering
\begin{equation}
  \label{eq:one}
  \mathcal{M}(s,\bm{\Delta}) = \frac{is}{2\pi}\int d\bm{x}_\perp
  e^{-i\bm{\Delta}\cdot \bm{x}_\perp}D(s,\bm{x}_\perp)~,
\end{equation}
where $s$ is the square of the center-of-mass energy, $\bm{\Delta}$ is the
momentum transfer, $\bm{x}_\perp$ is the impact parameter and all
spin variables have been omitted. For this model we take the simplest form that
we can use for the opacity
\begin{equation}
  \label{eq:opac}
D(s,\bm{x}_\perp) = 1-e^{-\Omega(s,\bm{x}_\perp)}~,
\end{equation}
with
\begin{equation}
  \label{eq:omDef}
  \Omega(s,\bm{x}_\perp)=\mathcal{S}(s)F(x_\perp)~,
\end{equation}
where $x_\perp \equiv |\bm{x}_\perp|$~.
The function $\mathcal{S}(s)$ is given by the complex symmetric expression, obtained from the high energy behavior
of quantum field theory \cite{chengWu}
\begin{equation}
  \label{eq:Sdef}
  \mathcal{S}(s)= \frac{s^c}{(\ln s)^{c'}} + \frac{u^c}{(\ln u)^{c'}}~,
\end{equation}
with $s$ and $u$ in units of $\mbox{GeV}^2$, where $u$ is the
third Mandelstam variable. In this Eq. (\ref{eq:Sdef}), $c$ and $c'$ are two 
dimensionless constants given below in Table 1. That they
are constants implies that the Pomeron is a fixed Regge cut rather than a
Regge pole. For the asymptotic behavior at high energy and modest momentum
transfers, we have to a good approximation
\begin{equation}
  \label{eq:uAp}
  \ln u = \ln s -i\pi~,
\end{equation}
so that
\begin{equation}
  \label{eq:S2}
    \mathcal{S}(s)= \frac{s^c}{(\ln s)^{c'}} + 
\frac{s^ce^{-i\pi c}}{(\ln s-i\pi)^{c'}}~.
\end{equation}
Because $F$ depends on $\bm{x}_\perp$ only through $x_\perp$, the Fourier transform in Eq. (\ref{eq:one}) simplifies to
\begin{equation}
  \label{eq:M2nd}
  \mathcal{M}(s,\Delta) = is\int_0^\infty dx_\perp\,x_\perp\,J_0(x_\perp\Delta)
\left[1-e^{-\mathcal{S}(s)F(x_\perp)}\right]~,
\end{equation}
where $\Delta \equiv |\bm{\Delta}|$.
The function $F(x_\perp)$ is taken to be related to the electromagnetic form factor
$G(t)$ of the proton, where $t= -\bm{\Delta}^2$ is the Mandelstam variable for
the square of the momentum transfer.  Specifically, $F(x_\perp)$ is defined as in
\cite{BSW79} via its Fourier transform $\tilde{F}(t)$ by
\begin{equation}
  \label{eq:FtildDef}
  \tilde{F}(t) = f[G(t)]^2\frac{a^2+t}{a^2-t}~,
\end{equation}
with 
\begin{equation}
  \label{eq:Gdef}
  G(t)=\frac{1}{(1- t/m_1^2)(1- t /m_2^2)}~.
\end{equation}
The remaining four parameters of the model, $f$, $a$, $m_1$ $\mbox{and}$ $m_2$, are given in Table 1.\\
The task is to study the asymptotic behavior of $\mathcal{M}$ for large $\ln s$
and modest momentum transfers. Before considering the asymptotic behavior we present a summary of
our predictions for the LHC energies.

 \begin{table}
    \centering
    \begin{tabular}{|rllrll|}\hline
$c$ & = & 0.167, & \;\;$c'$&=&0.748\\
$m_1$&=&0.577 GeV,&\;\;$m_2$&=&1.719 GeV\\
$a$&=&1.858 GeV,&\;\;$f$&=&6.971 GeV$^{-2}$\\\hline
    \end{tabular}
    \caption{\label{tab:table1} Parameters of the BSW model \cite{BSW03}.}
      \end{table}
\section{Predictions at LHC energies}
\label{sec2}
Two experiments are running at the nominal LHC energy $\sqrt{s} = 7\mbox{TeV}$,
TOTEM \cite{totem} and ATLAS-ALFA \cite{alfa} to measure p p elastic scattering, but so far 
only TOTEM has released preliminary data.
In view of a comparison with these experiments
we present here the predictions of BSW compared to TOTEM.
TOTEM forward slope $\mbox{B}^{T} = 20.1\pm 0.2\pm0.3$ GeV$^{-2}$ for $|t|$ (0.02-0.33)GeV$^2$, 
an extrapolation to t = 0 gives $\sigma_{tot}^{T} = 98.3 \pm 0.2\pm 2.8$ mb, our model gives
a continuous variation of the slope with $t$ (see \cite{BSW11}) but
an average slope over the previous t interval gives B = 19.4 GeV$^{-2}$, and 
$\sigma_{tot} = 92.7 \pm 0.8$ mb, also $\rho = 0.126 \pm 0.01$. 
Elastic cross section  $\sigma_{el}^{T} = 24.8\pm 0.2 \pm 1.2$ mb our prediction is  24.25 $\pm 0.3$mb, finally for the dip position
$|t_{dip}^{T}| = 0.53\pm 0.1\pm 0.1$ GeV$^2$, we obtain the same value. 

\begin{figure}[ht]
\hspace*{-150mm}
\begin{center}
  \epsfig{figure=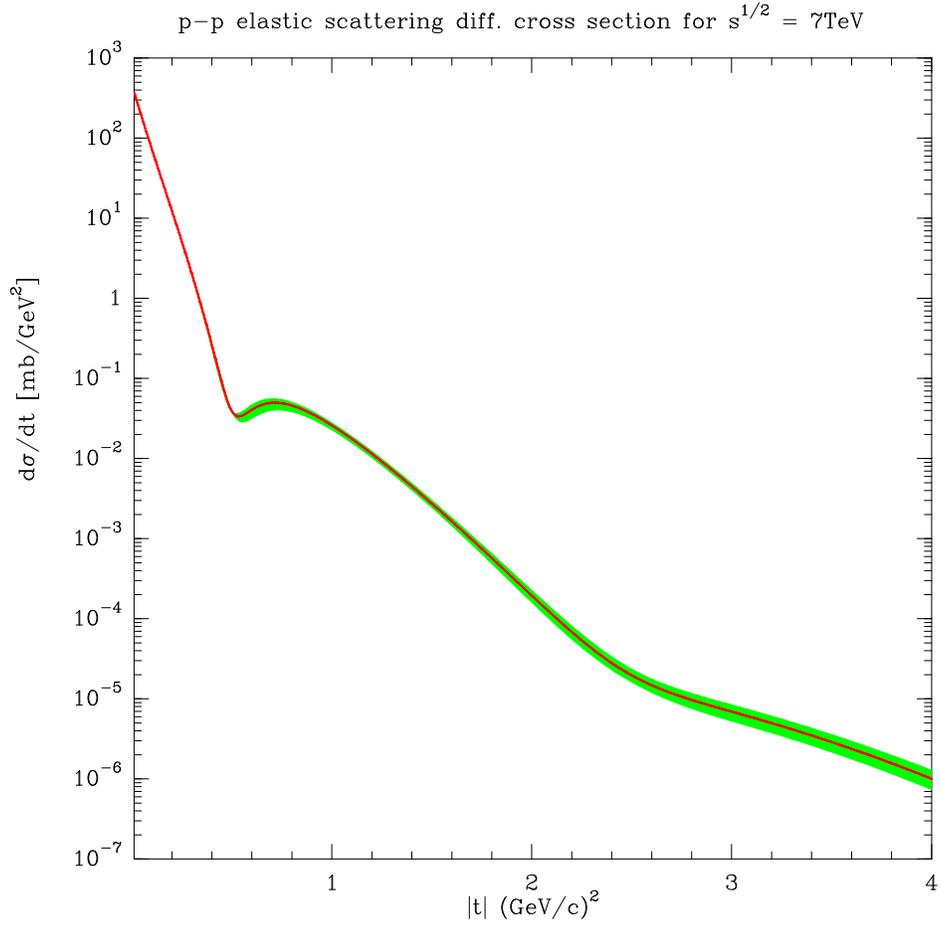,width=13.5cm}
\end{center}
  \vspace*{-20mm}
\caption{BSW prediction of the differential cross section versus $|t|$ for 
$\sqrt{s} = 7\mbox{TeV}$, uncertainties are calculated with a 68\% CL.}
\label{dsig7tev}
\vspace*{-1.5ex}
\end{figure}

The predicted BSW differential cross section is shown in Fig. \ref{dsig7tev} with uncertainties
calculated with a 68\% CL,\footnote{In the following all the BSW differential cross sections are calculated
 with a 68\% CL.} 
we cannot make an exact comparison with experiment since no final data are available, 
qualitatively we observe that the BSW differential cross section is above TOTEM 
at the second maximum by a factor around 1.7.

In view of a future experiment at $\sqrt{s} = 14\mbox{TeV}$ we give our predictions:
$\sigma_{tot} = 103.63 \pm 1.0$ mb, $\rho = 0.122 \pm 0.02$, the slope B near the forward direction
gives
20.15GeV$^{-2}$, $\sigma_{el}= 28.76 \pm 0.2 $ mb, and the elastic differential cross section
is shown in Fig. \ref{sig14} where $|t_{dip}| = 0.45$ GeV$^2$.
\section{The evaluation of $F(x_\perp)$ and its consequences}
\label{sec3}
The purpose of this section is to find the exact expression of $F(x_\perp)$, entering
in Eq.~(\ref{eq:M2nd}), in order to determine the most relevant region in $x_{\perp}$ for
the calculation of the asymptotic limit of $\mathcal{M}(s,\Delta)$, for large $s$.\\
Noting that $\tilde{F}$ depends only
on $\Delta^2$, the Fourier transform that defines $F$ simplifies to an
integral over one variable, so that we have
\begin{equation}
  \label{eq:Fsec}
  F(x_\perp)=\int_0^\infty d\Delta\,\Delta\,\tilde{F}(-\Delta^2)J_0(x_\perp\Delta)~,
\end{equation}
where $J_0$ denotes the Bessel function of zero order. From
Eq. (\ref{eq:FtildDef}), we have explicitely
\begin{equation}
  \label{eq:Ftild}
  \tilde{F}(-\Delta^2)= f\,
\frac{1}{(1+ \Delta^2/m_1^2)^{2}(1+ \Delta^2/m_2^2)^{2}}
\frac{a^2-\Delta^2}{a^2+\Delta^2}~,
\end{equation}
which is a rational fraction, symmetric in $m_1, m_2$, whose decomposition into simple elements, allows the direct
calculation of $F(x_\perp)$. As expected, the final result can be expressed in terms of modified Bessel 
functions $K_0$ and $K_1$. We have the decomposition
\begin{equation}
  \label{eq:Fpole}
  F(x_\perp)=F_1(x_\perp)+F_2(x_\perp)+F_3(x_\perp)~,
\end{equation}
where
\begin{eqnarray}
  \label{eq:R1K}
F_1(x_\perp)&=&\frac{f m_1^4m_2^4}{2(m_2^2-m_1^2)^3}
\bigg\{\frac{m_2^2-m_1^2}{m_1^2}\,\frac{a^2+m_1^2}{a^2-m_1^2}
(m_1x_\perp)K_1(m_1x_\perp)\nonumber\\
&&-4\left[\frac{a^2+m_1^2}{a^2-m_1^2}+
\frac{(m_2^2-m_1^2)a^2}{(a^2-m_1^2)^2}\right]K_0(m_1x_\perp)\bigg\}\\
F_2(x_\perp)&=&\frac{f m_1^4m_2^4}{2(m_2^2-m_1^2)^3}
\bigg\{\frac{m_2^2-m_1^2}{m_2^2}\,\frac{a^2+m_2^2}{a^2-m_2^2}
(m_2x_\perp)K_1(m_2x_\perp)\nonumber \\ &&+4\left[\frac{a^2+m_2^2}{a^2-m_2^2}-
\frac{(m_2^2-m_1^2)a^2}{(a^2-m_2^2)^2}\right]K_0(m_2x_\perp)\bigg\} \label{eq:R2K}\\
F_3(x_\perp)&=&\frac{2fm_1^4m_2^4a^2}{(a^2-m_1^2)^2(a^2-m_2^2)^2}K_0(ax_\perp)~,
\label{eq:R3K}
\end{eqnarray}
which is clearly symmetric in $m_1, m_2$.
The arguments of the modified Bessel functions $K_1$ and $K_0$ are $m_1x_\perp$ for $F_1$, while $m_2x_\perp$ for $F_2$ and
$ax_\perp$ for $F_3$. The function $F$ is real, positive, bounded, and monotonically
decreasing toward 0 as $x_\perp$ increases without bound. Because $\text{Re}
\mathcal{S}$ is large and positive for large values of
$\ln s$, the only contribution
to $\mathcal{M}$ comes from $x_\perp$ fairly near $\ln s$, so that the large
argument asymptotic expressions for the modified Bessel functions are
applicable. Since as seen from Table 1, $m_1< m_2 < a$, $F_2$ and $F_3$ are exponentially smaller than $F_1$,
only $F_1$ contributes to the asymptotic behavior of $\mathcal{M}(s,\Delta)$. For the moment we keep the exact expression for $F(x_\perp)$, but we change
variables to emphasize the role of $F_1$ as follows. Let us define two dimensionless variables
\begin{eqnarray}
  \label{eq:var}
  x=m_1x_\perp,\\
\alpha =\Delta/m_1.
\end{eqnarray}

With this change of variables, Eq. (\ref{eq:M2nd}) becomes
\begin{equation}
  \label{eq:M6}
  \mathcal{M}(s,\Delta) = \frac{is}{m_1^2}\int_0^\infty dx\,x\,J_0(\alpha x)
  \left[1-e^{-\mathcal{S}(s)\hat{F}(x)}\right]~,
\end{equation}
where $\hat{F}(x)= \hat{F}_1(x)+\hat{F}_2(x)+\hat{F}_3(x)$ and
 $\hat{F}_j(x)\equiv F_j(x/m_1)=F_j(x_\perp)$, for $j=1,2,3$.
We note that for $x\gg1$ the asymptotic
representation of $\hat{F}$, implied by Eqs.~(\ref{eq:R1K}) and (\ref{eq:var})
and the asymptotic representation of the modified Bessel function \cite{bII}, is
\begin{eqnarray}
  \hat{F}(x)& \sim &\frac{f m_1^2m_2^4(a^2+m_1^2)}{2(m_2^2-m_1^2)^2(a^2-m_1^2)}
x\,K_1(x) \\ &\sim &b  \sqrt{x}\,e^{-x},\label{eq:Fa}
\end{eqnarray}
where $b$ is the real coefficient
\begin{equation}
  \label{eq:bDef}
  b= \frac{f m_1^2m_2^4(a^2+m_1^2)}{2(m_2^2-m_1^2)^2(a^2-m_1^2)}\sqrt{\pi/2}. 
\end{equation}

\section{High-energy asymptotic behavior}
\label{sec4}
Some additional approximations make it possible to obtain an
 expression of the scattering amplitude in the high energy limit.
The integrand of the integral in Eq. (\ref{eq:M2nd}) is complex.
From the definition of $\mathcal{S}(s)$ in Eq. (\ref{eq:Sdef}) and the explicit expression in Eq. (\ref{eq:S2}), we notice that
for sufficiently high values of $s$, namely
\begin{equation}
  \label{eq:lns}
  \ln s \gg \pi~,
\end{equation}
the second denominator in Eq. (\ref{eq:S2}) can be approximated by the first denominator, so that
\begin{equation}
  \label{eq:Sarg}
  \mathcal{S}(s) \sim \frac{2s^c}{(\ln s)^{c'}}\cos\left(\frac{\pi c}{2}\right)
e^{-\frac{1}{2}i\pi c}~.
\end{equation}
Thus the phase of $S(s)$ is found to be a constant, namely, $-\frac{1}{2}\pi c \approx
-0.26$ and when $\ln s$ is large one has
\begin{equation}
  \label{eq:ss}
  \text{Re} \mathcal{S}(s) > \text{Im} \mathcal{S} \gg 1.
\end{equation}
The function $\hat{F}(x)$ is positive for all real $x$ and decreasing
exponentially with large $x$ as stated in
Eq.(\ref{eq:Fa}). Because of these properties of $\hat{F}(x)$ and the large value of
$\text{Re}\mathcal{S}$, the opacity is essentially 1 for values of $x$, well below a
transition value, and drops exponentially to zero as $x$ increases well above this
transition value. From this, it follows that the asymptotic representation of
the scattering amplitude can be obtained by replacing $\hat{F}(x)$ by
$\hat{F}(z_0)e^{-(x-z_0)}$, where $z_0$ is a complex transition value chosen in such a way to
make $\mathcal{S}(s)\hat{F}(z_0)$ real and of order 1.\\
By means of a translation in the complex plane from $x$ to $x' = x-z_0$, the
scattering amplitude of Eq.~(\ref{eq:M6}) becomes asymptotically
\begin{eqnarray}
  \label{eq:M5}
  \mathcal{M}(s,\Delta)&\sim&\frac{is}{m_1^2}\int_{-z_0}^\infty dx'(x'+z_0)
  J_0[\alpha(x'+z_0)]\left[1-e^{-\mathcal{S}(s)\hat{F}(z_0)e^{-x'}}\right]\\
&=&\frac{is}{m_1^2}\left[\int_{-z_0}^0 dx'(x'+z_0)
  J_0[\alpha(x'+z_0)]+A(s,z_0)\right] \label{eq:M3b} \\ &=&
\frac{is}{m_1^2}\left[\frac{z_0}{\alpha}J_1(\alpha z_0)+A(s,z_0)\right]~,\label{eq:M3c}
\end{eqnarray}
where we define
\begin{eqnarray}
  A(s,z_0)&=&\int_0^\infty dx'(x'+z_0)J_0[\alpha(x'+z_0)]
\left[1-e^{-\mathcal{S}(s)\hat{F}(z_0)e^{-x'}}\right]\nonumber\\
&& -\int_{-z_0}^0 dx'(x'+z_0)J_0[\alpha(x'+z_0)]
e^{-\mathcal{S}(s)\hat{F}(z_0)e^{-x'}}  \label{eq:Adef}.
\end{eqnarray}
It remains to determine $z_0$ and then to evaluate the asymptotic representation of $A(s,z_0)$.
 In order to find $z_0$ we use the following  relation derived in Appendix A (see Eq.~(\ref{eq:ch}))
\begin{equation}
  \label{eq:ch1}
  0=\int_0^\infty dx 
\left[1-e^{\displaystyle{-e^{-\gamma} e^{-x}}}\right] 
-\int_{-\infty}^0 dx\,e^{\displaystyle{-e^{-\gamma} e^{-x}}}~,
\end{equation}
which suggests defining $z_0$ as solution of the equation
\begin{equation}
  \label{eq:z0Def}
  \mathcal{S}(s)\hat{F}(z_0)=e^{-\gamma}~,
\end{equation}
where $\gamma \approx 0.5772$ is the Euler's constant.  While $z_0$, as a function of $s$, is
best obtained by solving Eq.~(\ref{eq:z0Def}) numerically, one can see its
approximate value by using the asymptotic representation for $\hat{F}$, so that
\begin{equation}
  \label{eq:z0ap}
 \mathcal{S}(s)b\sqrt{z_0}\,e^{-z_0}\approx e^{-\gamma}.
\end{equation}
By taking logarithms one finds
\begin{equation}
  \label{eq:z0ap2}
  z_0\approx \ln[b\sqrt{z_0}\mathcal{S}(s)]+\gamma,
\end{equation}
showing how $\text{Re}\,z_0$ grows with $\text{Re}\,\ln[\mathcal{S}(s)]$ and how
$\text{Im}\,z_0 = \arg(\mathcal{S}(s))$  approaches $-0.26$, as $\ln s$
becomes very large (see Fig. 1). Then the fact that $\text{Re}\,z_0$ is large can be used twice. First, in Eq.~(\ref{eq:Adef}), one can safely
 replace the lower limit $-z_0$ in one of the
integrals by $-\infty$, so that
\begin{eqnarray}
  A(s,z_0)&\sim & \int_0^\infty dx(x+z_0)J_0[\alpha(x+z_0)]
\left[1-e^{-e^{-(x+\gamma)}}\right] \nonumber\\
&& -\int_{-\infty}^0 dx(x+z_0)J_0[\alpha(x+z_0)]
e^{-e^{-(x+\gamma)}}. \label{eq:Az0}
\end{eqnarray}
Secondly, since the important regions of integration are where $x$ is of the order of $1$, to determine the leading order behavior of the scattering amplitude it suffices to replace the factors $(x +z_0)$ by $z_0$, so Eq.~(\ref{eq:Az0}) reduces to
\begin{eqnarray}
  A(s,z_0)&\sim & z_0\bigg\{\int_0^\infty dx\,J_0[\alpha(x+z_0)]
\left[1-e^{-e^{-(x+\gamma)}}\right] \nonumber\\
&& -\int_{-\infty}^0 dx\,J_0[\alpha(x+z_0)]
e^{-e^{-(x+\gamma)}}\bigg\}. \label{eq:Az1}
\end{eqnarray}

To determine the asymptotic behavior of $A(s,z_0)$, we have to study the Bessel function which depends upon the parameter $\alpha$, and there are two regions to consider.\\

{\bf\small{i) Small $\alpha$} region}\\
By assuming that $\alpha$ is small, but not $\alpha z_0$, we keep only the first three terms
of the Taylor series to obtain:
\begin{equation}
  \label{eq:tay2}
  J_0[\alpha(z_0+x)] = J_0(\alpha z_0)-\alpha x J_1(\alpha z_0)
-\frac{(\alpha x)^2}{2}J_0(\alpha z_0)
+ \text{higher order terms}.
\end{equation}
By substituting Eq.~(\ref{eq:tay2}) into Eq.~(\ref{eq:Az1}), one obtains after
some integration by parts
\begin{eqnarray}
  \label{eq:A2}
  A(s,z_0)&\sim&z_0\bigg[\mathcal{I}_1(e^{-\gamma})J_0(\alpha z_0) -
\frac{\alpha}{2}\mathcal{I}_2(e^{-\gamma}) J_1(\alpha z_0) \nonumber \\&&
  -\frac{\alpha^2}{6}\mathcal{I}_3(e^{-\gamma})
J_0(\alpha z_0)\bigg]\nonumber\\
&=&-z_0\left[\frac{\pi^2\alpha}{12}J_1(\alpha
  z_0)+\frac{\alpha^2}{3}\zeta(3) J_0(\alpha z_0)
\right],\label{eq:A2b}
\end{eqnarray}
where the $\mathcal{I}_n$ are defined in Appendix A and $\zeta$ is the Riemann Zeta function, and therefore $\zeta(3) \approx1.2021$.\\

{\bf\small{ii) Large $\alpha$} region}\\
In this region the large argument asymptotic expansion of the Bessel function
\cite{bII} allows one to write, for large $|z_0|$ and the contributing values of
$x$ which are $O(1)$,
\begin{eqnarray}
  J_0[\alpha(z_0+x)]&= &\sqrt{\frac{2}{\pi z_0}}\cos[\alpha(z_0+x-\pi/4)] + 
\text{higher order terms}\nonumber
\\ &=&\frac{1}{\sqrt{2\pi z_0}}\left[e^{i\alpha(z_0-\pi/4)}e^{i\alpha x}+
e^{-i\alpha(z_0-\pi/4)}e^{-i\alpha x}\right] + . . . . \label{eq:J0}
\end{eqnarray}
By substituting Eq.~(\ref{eq:J0}) into Eq.~(\ref{eq:Az1}), this yields for this region
of $\alpha$
\begin{eqnarray}
  A(s,z_0)&\sim&\sqrt{\frac{z_0}{2\pi}}\left[e^{i(\alpha z_0-\pi/4)}
\mathcal{J}(i\alpha) +e^{-i(\alpha z_0-\pi/4)}\mathcal{J}(-i\alpha)\right] \\ 
&=& z_0 \sqrt{\frac{2}{\pi z_0}}\left[\cos(\alpha z_0-\pi/4)\text{Re}
\mathcal{J}(i\alpha)- \sin(\alpha z_0-\pi/4)\text{Im}\mathcal{J}(i\alpha)\right]\nonumber,
\end{eqnarray}
where it follows from Eq.~(\ref{eq:A11}) of Appendix A that
\begin{equation}
  \label{eq:JJ}
  \mathcal{J}(i\alpha)=\frac{i}{\alpha}- e^{-i\gamma\alpha}\Gamma(-i\alpha)~. 
\end{equation}
Comparison with Eq.~(\ref{eq:J0}) and the similar expression for $J_1$ shows that
to leading order, this result can be expressed in terms of Bessel functions
\begin{eqnarray}
&&  A(s,z_0)\sim z_0\left[J_0(\alpha z_0)\text{Re}\mathcal{J}(i\alpha)- J_1(\alpha
    z_0)\text{Im}\mathcal{J}(i\alpha)\right]     \nonumber \\
&&= -z_0\!\!\left[\!\left(\frac{1}{\alpha}+
\text{Im}\left[e^{i\gamma \alpha}\Gamma(i\alpha)\right]\right)J_1(\alpha z_0)
\!\!+\!\!\left(\text{Re}\left[e^{i\gamma \alpha}\Gamma(i\alpha)\right]\right)J_0(\alpha
z_0)\!\right].
\label{eq:Alead2}
    \end{eqnarray}
{\bf\small{iii) Uniform approximation}}\\
Expanding the Gamma function for small $\alpha$, one obtains Eq.~(\ref{eq:A2b}),
showing that Eq.~(\ref{eq:Alead2}) gives a uniform approximation including both
regions of the parameter $\alpha$. Then from Eq.~(\ref{eq:M3c}) it follows that
for $0 \le \alpha \ll |z_0|$, the asymptotic representation of the scattering
amplitude is
\begin{equation}
  \label{eq:M4}
  \mathcal{M}(s,\Delta) \sim -\frac{isz_0}{m_1^2}
\left\{\left[\text{Im}(e^{i\gamma \alpha}\Gamma(i\alpha))\right]J_1(\alpha z_0)
+    \left[\text{Re}(e^{i\gamma \alpha}\Gamma(i\alpha))\right]J_0(\alpha z_0)\right\}~.
\end{equation}

\section{Numerical results}
\label{sec5}
\begin{figure}[htbp]
\begin{center}
\hspace*{-80mm}
  \begin{minipage}{6.5cm}
  \epsfig{figure=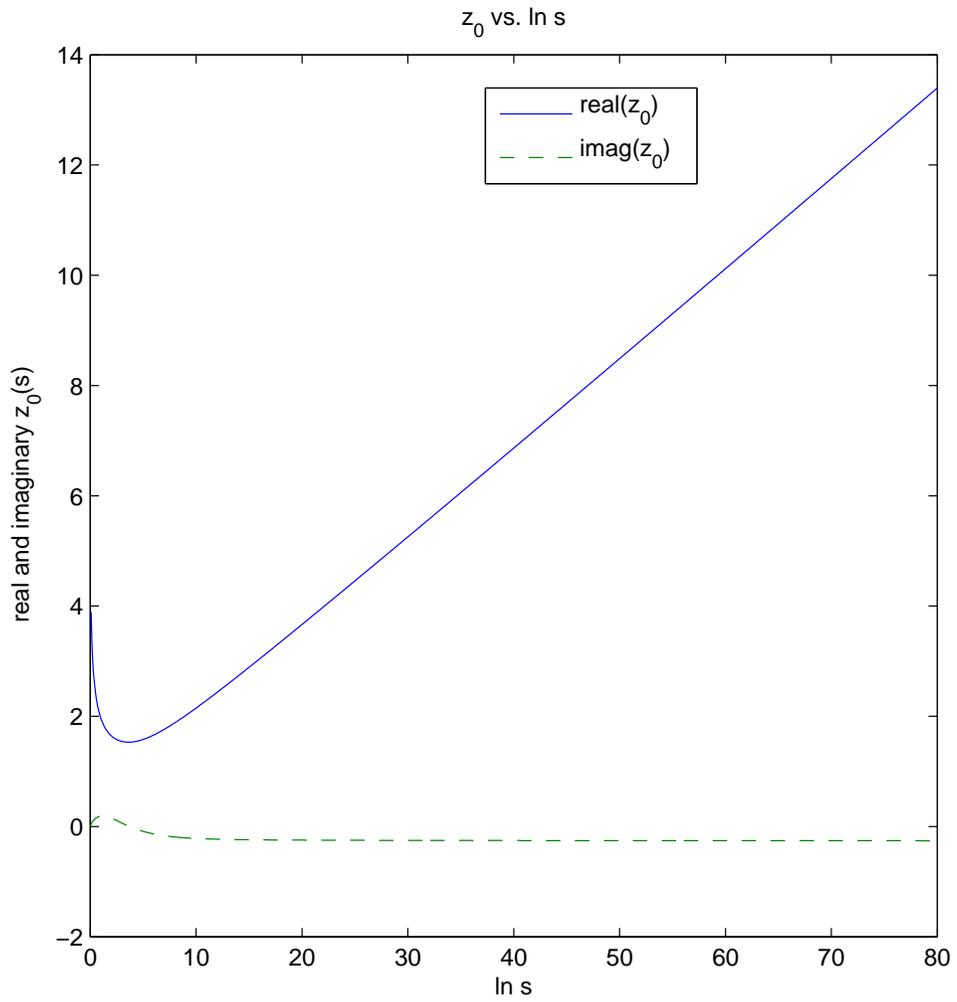,width=14.5cm}
  \end{minipage}
\end{center}
  \vspace*{-10mm}
\caption{
The real and imaginary parts of $z_0(s)$ as a function of the energy, obtained by
solving numerically Eq.~(\ref{eq:z0Def}).}
\label{z0}
\end{figure}

In this section we present some numerical results to illustrate the 
asymptotic formulas
obtained on the physical quantities of interest for different energy values. 
First let us come back to the determination of the 
complex 
parameter $z_0(s)$ which plays an essential role in the solution of the
high energy asymptotic behavior of the scattering amplitude. As we said earlier, 
$z_0$ is
obtained by solving numerically Eq.~(\ref{eq:z0Def}) and the results are shown on 
Fig. \ref{z0}, for 
$\text{Re}z_0(s)$ and $\text{Im}z_0(s)$, versus $\ln{(s)}$. As expected, 
$\text{Re}z_0(s)$ rises rapidly with $\ln{(s)}$,
whereas $\text{Im}z_0(s)$ remains small and almost energy independent. 
It is worth noting that the
asymptotic regime requires the validity of Eq.~(\ref{eq:lns}), for example for
 $\ln{(s)} = 10\pi$,
corresponding to the center-of energy $\sqrt{s}$ of about 6000TeV.

\begin{figure}[htbp]
\begin{center}
\hspace*{-50mm}
  \begin{minipage}{6.5cm}
  \epsfig{figure=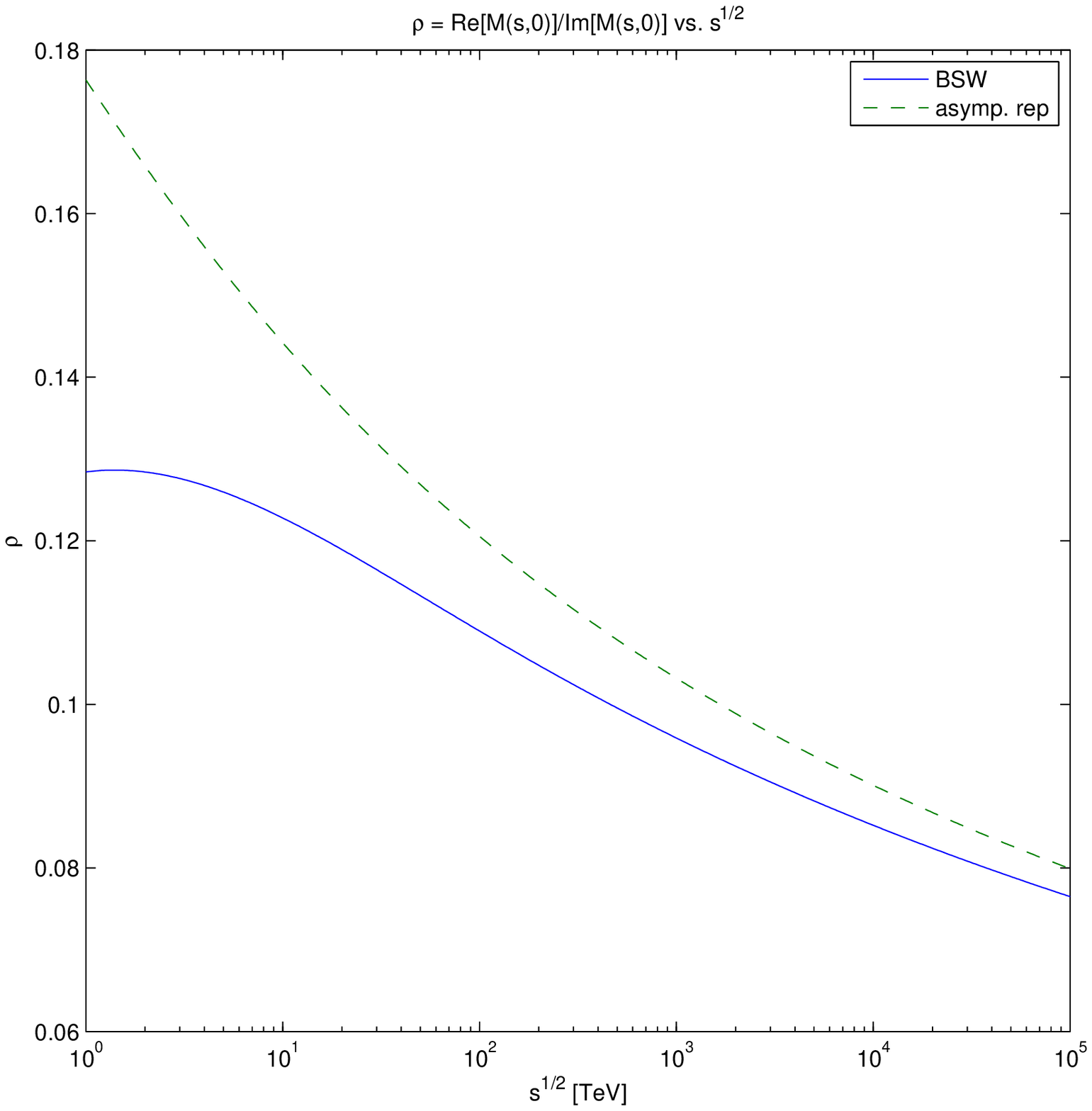,width=10.5cm}
    \epsfig{figure=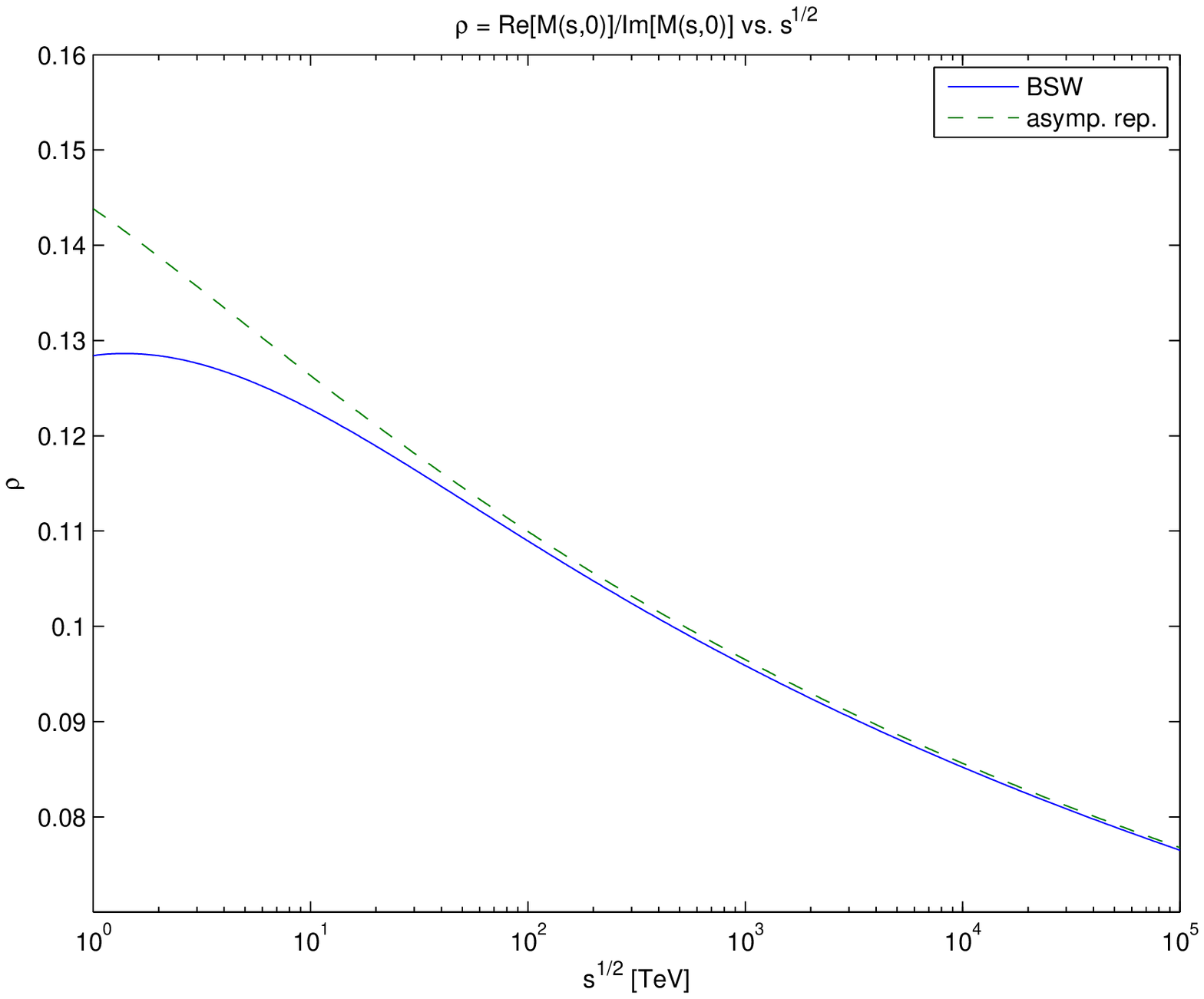,width=10.5cm}
  \end{minipage}
\end{center}
\caption{
The ratio of the real to the imaginary parts of the forward amplitude versus 
$\sqrt{s}$. 
Top using Eq. (\ref{eq:fscatt}) and
bottom using Eq. (\ref{eq:fscatt3}) dashed curves, BSW solid curves.}
\label{rho}
\vspace*{-1.5ex}
\end{figure}

\begin{figure}[htbp]
\begin{center}
\hspace*{-50mm}
  \begin{minipage}{6.5cm}
  \epsfig{figure=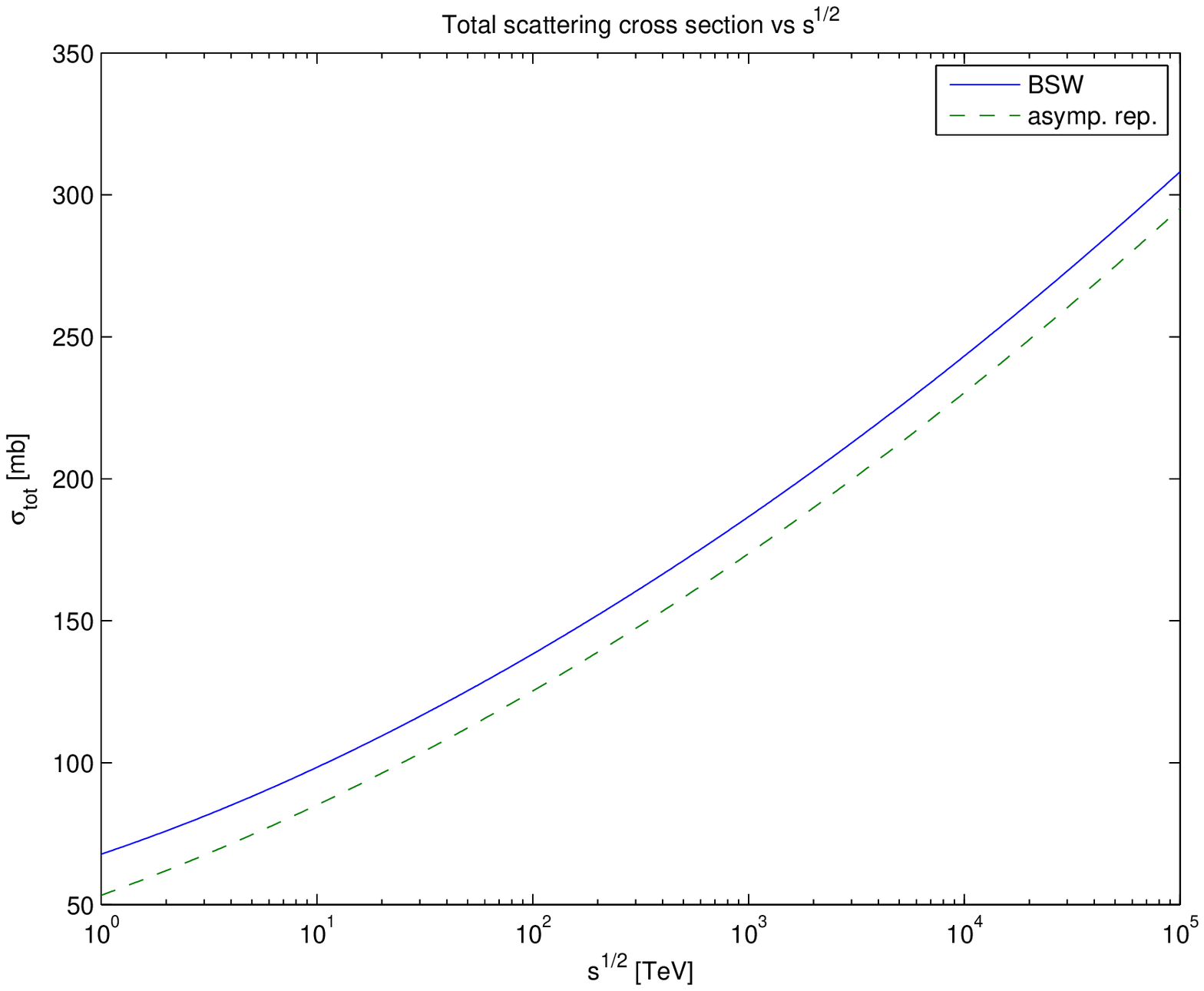,width=10.5cm}
    \epsfig{figure=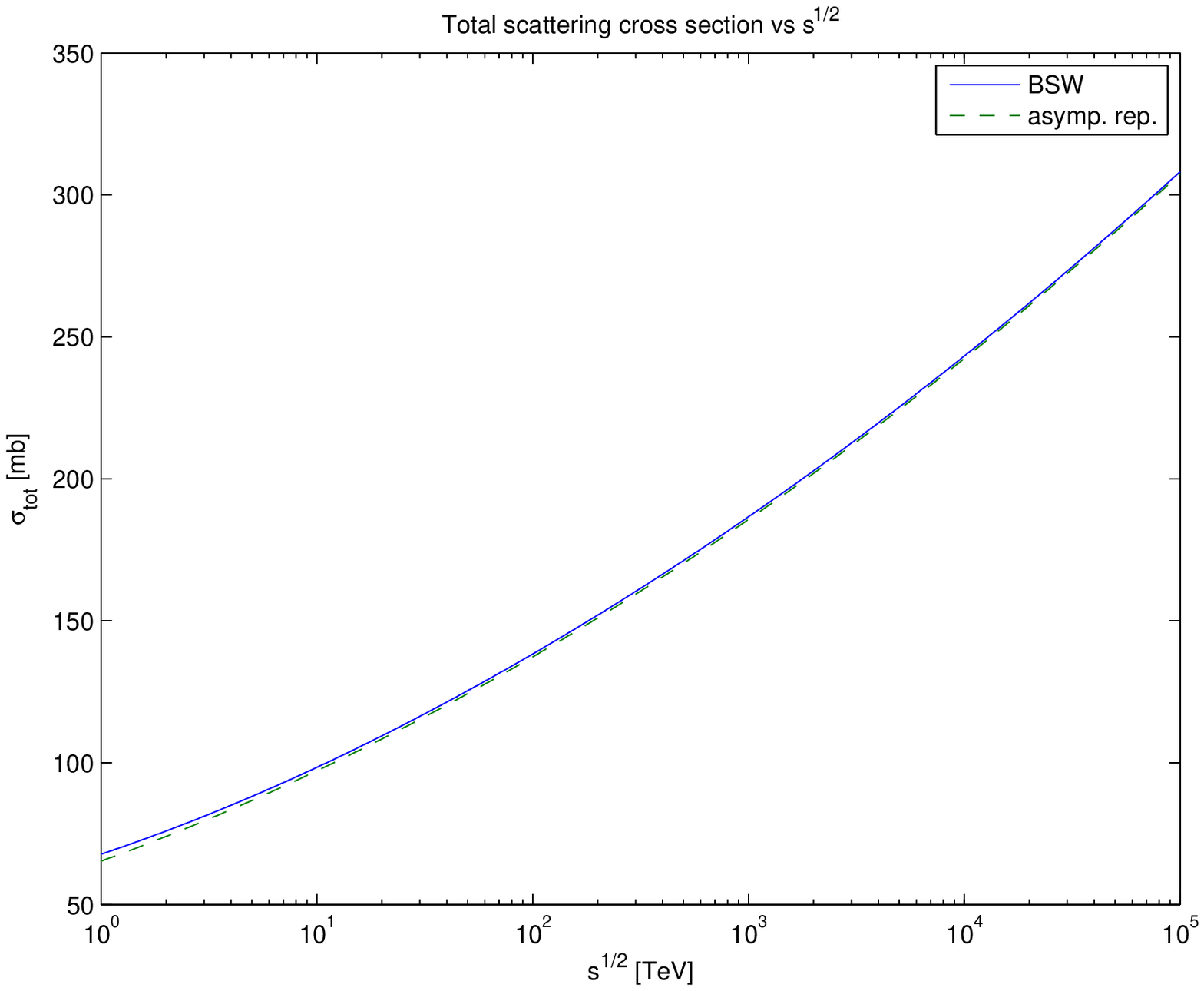,width=10.5cm}
  \end{minipage}
\end{center}
  \vspace*{-10mm}
\caption{
The total cross section $\sigma_{tot}$ versus $\sqrt{s}$.  Top using 
Eq. (\ref{eq:fscatt}) and bottom using Eq. (\ref{eq:fscatt3}) dashed curves,
BSW solid curves.}
\label{tot}
\vspace*{-1.5ex}
\end{figure}

\begin{figure}[htbp]
\begin{center}
\hspace*{-80mm}
  \begin{minipage}{6.5cm}
  \epsfig{figure=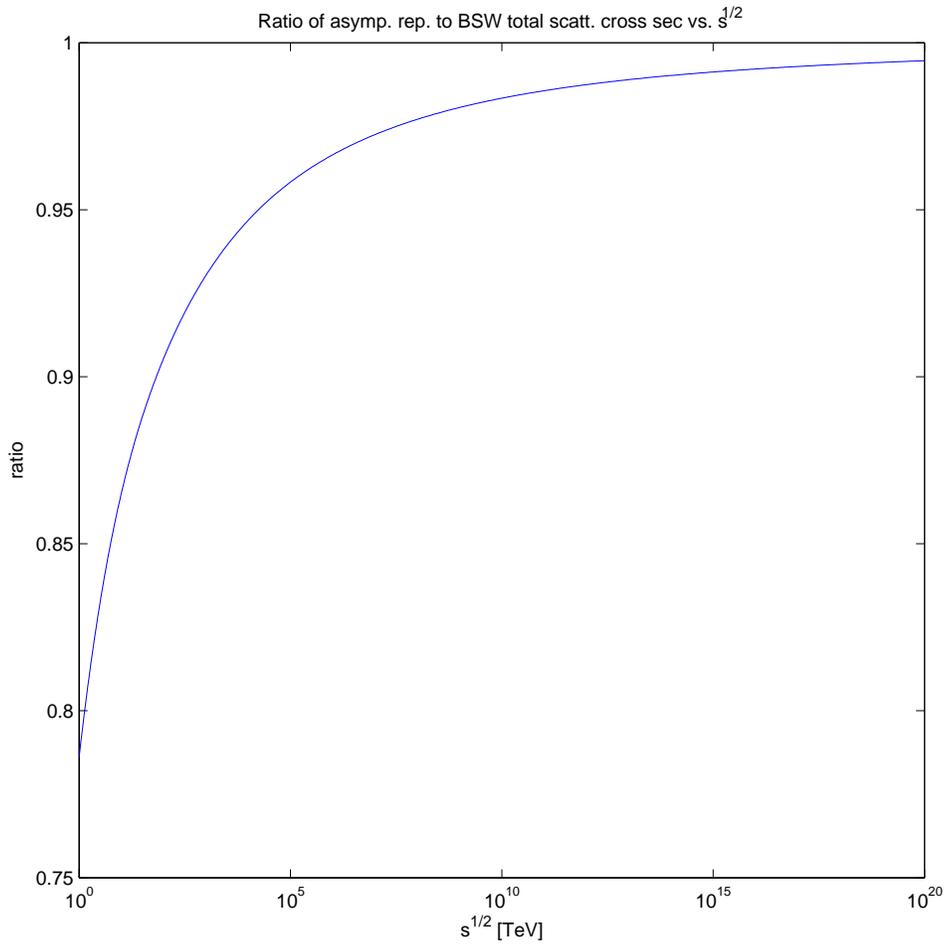,width=14.5cm}
  \end{minipage}
\end{center}
  \vspace*{-10mm}
\caption{
The ratio of the leading order of the asymptotic representation to the exact BSW result, versus the energy.}
\label{ratio}
\vspace*{-1.5ex}
\end{figure}

\begin{figure}[htbp]
\begin{center}
\hspace*{-80mm}
  \begin{minipage}{6.5cm}
  \epsfig{figure=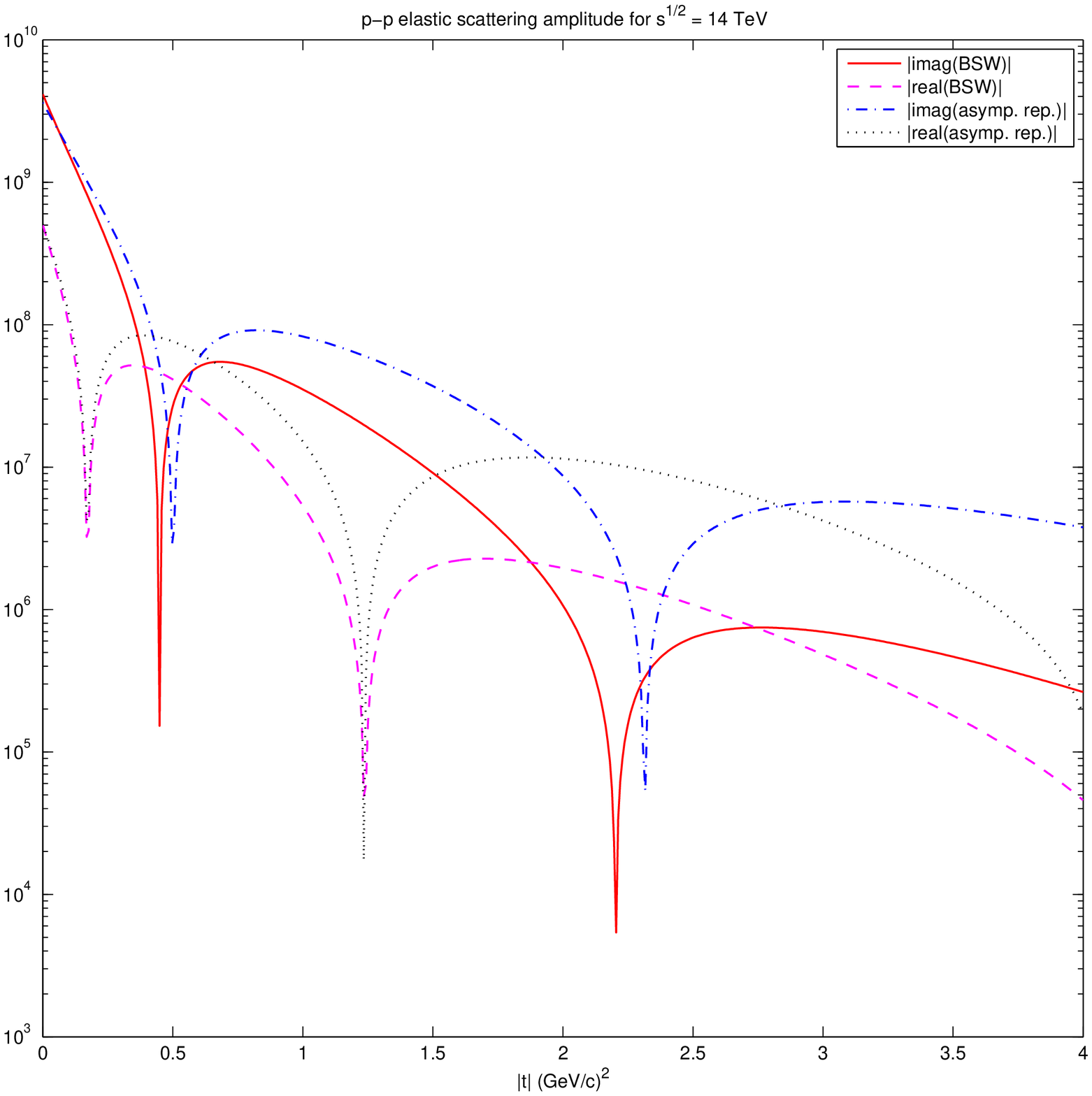,width=14.5cm}
  \end{minipage}
\end{center}
  \vspace*{-10mm}
\caption{
The absolute value of the real and imaginary parts of the elastic scattering 
amplitude, 
as a function of $|t|$ for $\sqrt{s} = 14\text{TeV}$, for the exact BSW result 
(Real: dashed, Im: solid)
and the asymptotic representation (Real: dotted, Im: dash-dotted).}
\label{amp14}
\vspace*{-1.5ex}
\end{figure}

\begin{figure}[t]
\hspace*{-200mm}
\begin{center}
  \epsfig{figure=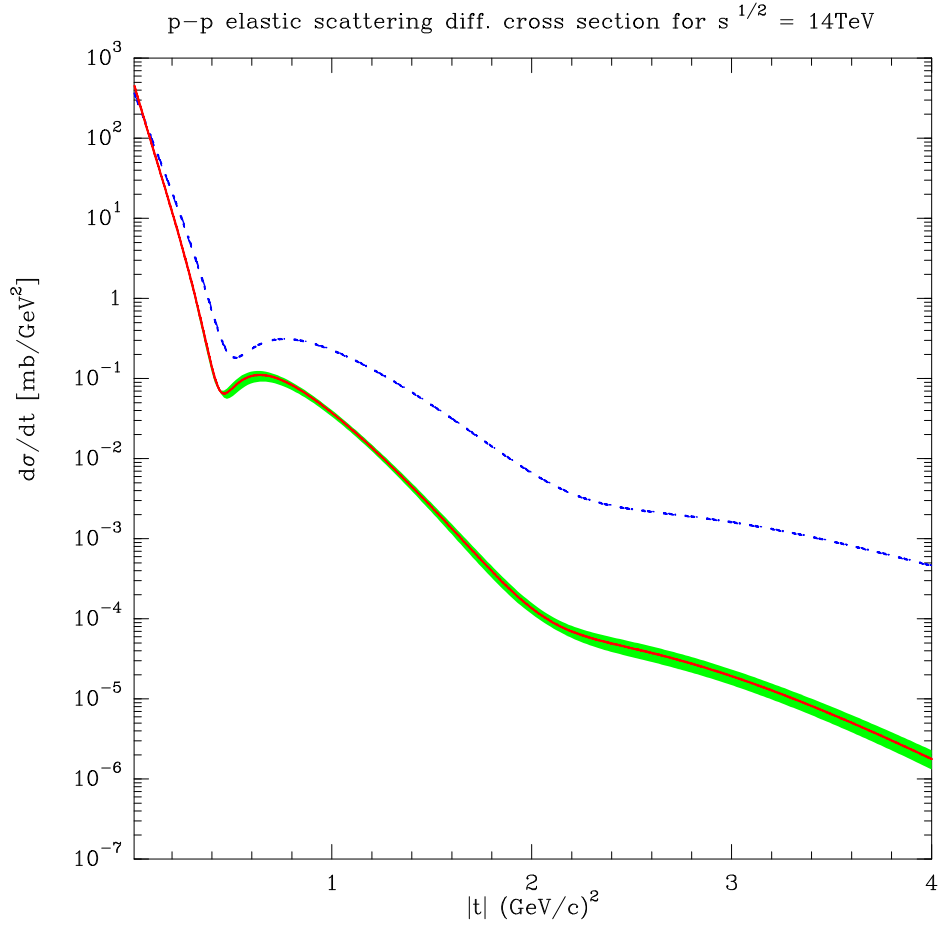,width=13.5cm}
\end{center}
  \vspace*{-20mm}
\caption{
The elastic differential cross section versus $|t|$ for $\sqrt{s} = 14\mbox{TeV}$
calculated using Eq.~(\ref{eq:sig1}) dashed curve, BSW solid curve.}
\label{sig14}
\end{figure}

\begin{figure}[htbp]
\begin{center}
\hspace*{-80mm}
  \begin{minipage}{6.5cm}
  \epsfig{figure=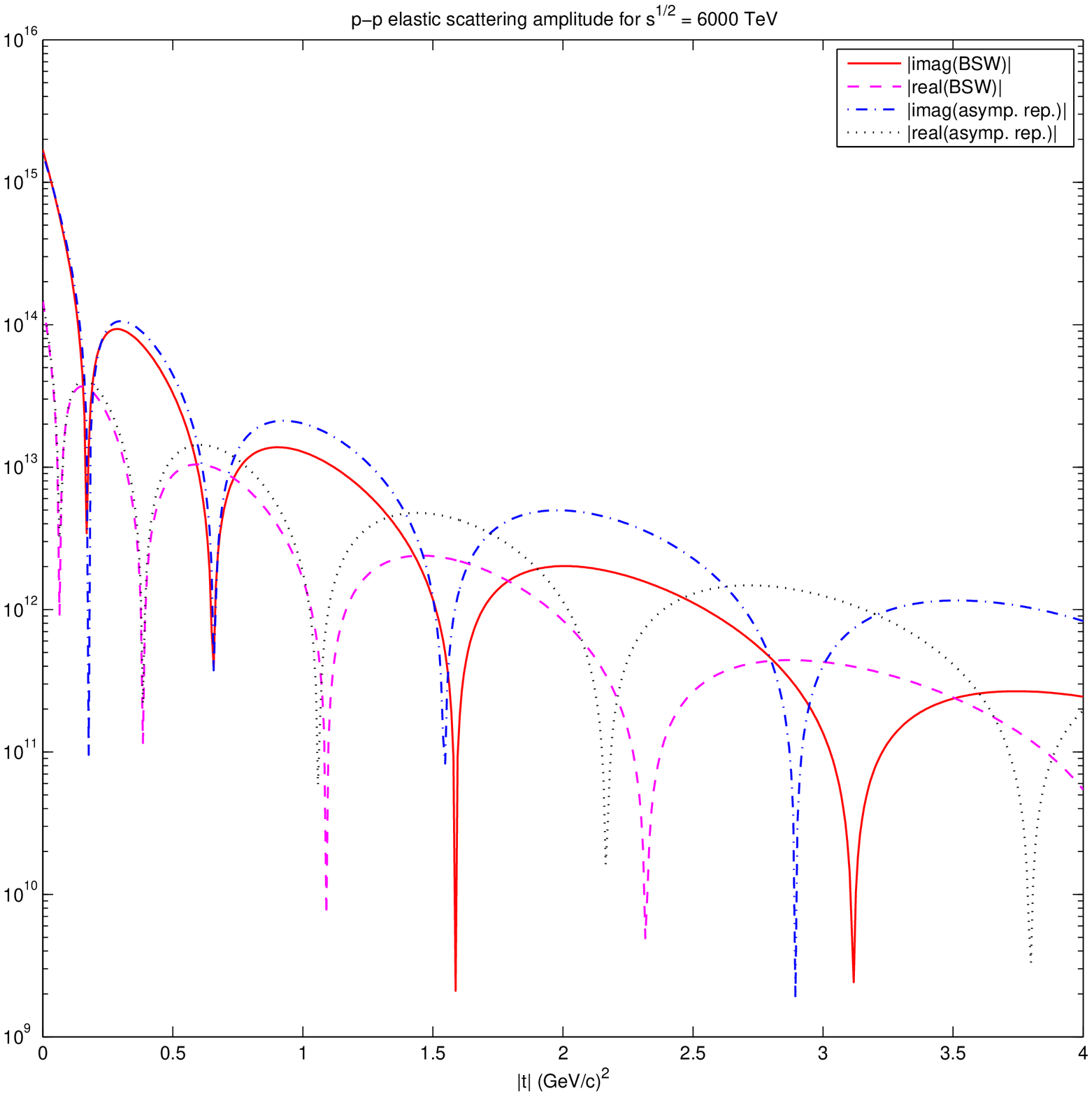,width=14.5cm}
  \end{minipage}
\end{center}
  \vspace*{-10mm}
\caption{The absolute value of the real and imaginary parts of the elastic scattering 
amplitude, as a function of $|t|$ for $\sqrt{s} = 6000\text{TeV}$, , for the exact BSW result 
(Real: dashed, Im: solid)
and the asymptotic representation (Real: dotted, Im: dash-dotted).}
\label{amp6000}
\vspace*{-1.5ex}
\end{figure}

\begin{figure}[ht]
\hspace*{-150mm}
\begin{center}
  \epsfig{figure=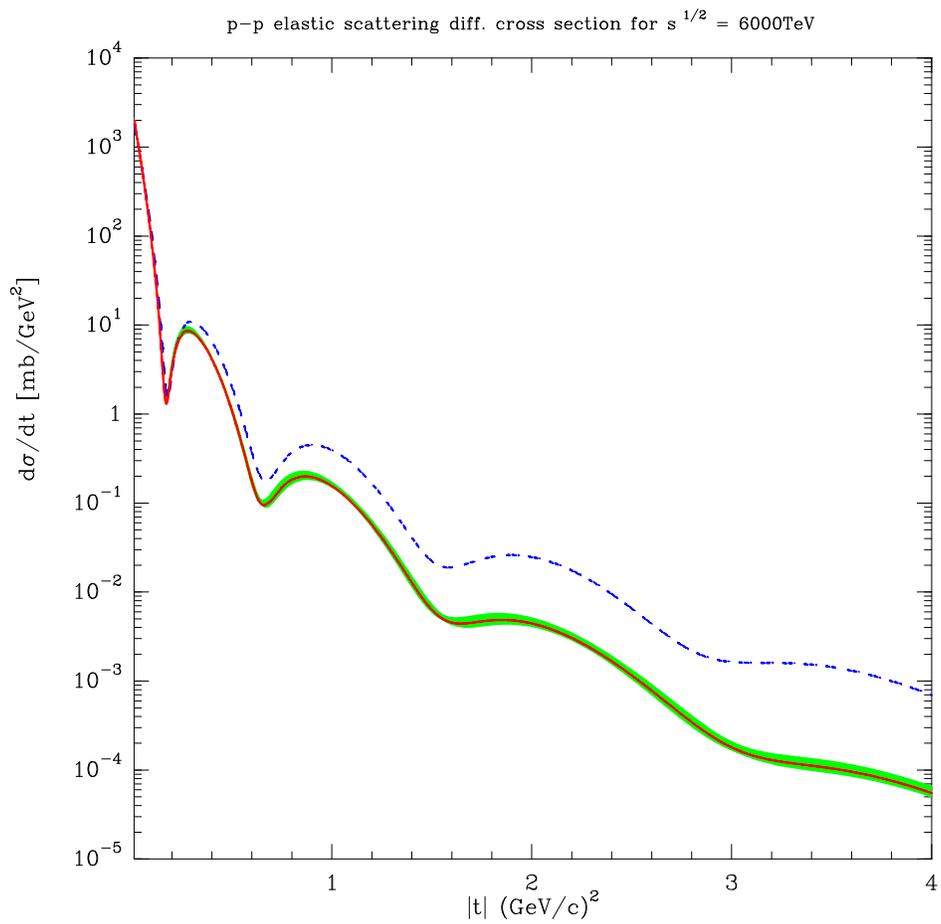,width=13.5cm}
\end{center}
  \vspace*{-20mm}
\caption{
The elastic differential cross section calculated using Eq.~(\ref{eq:sig1})
dashed curve, BSW solid curve.}
\label{sig6000}
\vspace*{-1.5ex}
\end{figure}

The asymptotic representation of the forward scattering amplitude is obtained
from Eq.~(\ref{eq:M4}) by taking the limit as $\alpha \rightarrow 0$
\begin{equation}
  \label{eq:fscatt}
  \mathcal{M}(s,0) \sim \frac{isz_0^2}{2m_1^2}~.
\end{equation}
From this formula we calculate the ratio of the real to the imaginary parts of the forward amplitude defined as 
\begin{equation}
  \label{eq:rho}
\rho(s) = \frac{\text{Re}\mathcal{M}(s,0)}{\text{Im}\mathcal{M}(s,0)}~.
\end{equation}
In Fig. \ref{rho} (top) we display this result compared to the exact BSW result. We see that $\rho(s)$
decreases for increasing energy, in agreement with the expectation that $\rho(s) \to 0$, when $s$ goes to infinity.\\
The total cross section is obtained from the optical theorem as follows, 
\begin{equation}
  \sigma_{\text{tot}}=\frac{4\pi}{s}\text{Im}\mathcal{M}(s,0),
\end{equation}
and we recall that $\mathcal{M}(s,0)$ is dimensionless. It
 is plotted in Fig \ref{tot} (top) compared to the exact BSW result.\\
 In Figs. \ref{rho}, \ref{tot} (top), a gap can be noticed between the asymptotic 
representation and
the BSW model. The gaps extend to the end of the plotted energy range of
$s^{1/2}=10^5$ TeV, where $\ln{(s)} \approx 37$. To show that the asymptotic
representation actually approaches BSW at sufficiently large values of
$\ln s$, we carried out the first three terms of an asymptotic expansion to
obtain
\begin{equation}
 \label{eq:fscatt3}
 \mathcal{M}(s,0) \sim \frac{is}{2m_1^2}
 \left(z_0^2+\frac{\pi^2}{6}\right)+O(1/z_0)~.
\end{equation}
With this expression in place of (42), the gaps largely close, as seen in Figs. \ref{rho}, \ref{tot}
(bottom). This shows indeed that the
gaps visible in Figs. \ref{rho}, \ref{tot} (top) are due to the neglect of non-leading terms in the 
asymptotic
representation. In Fig. \ref{ratio} we display the ratio of the leading order of the asymptotic 
representation to the exact BSW result, which
goes to 1 for very, very large $s$, as expected.
 Now let us move from the forward direction to look at the behavior of the real 
and imaginary parts
 of the scattering amplitude as functions of $t$. For $\sqrt{s} = 14\text{TeV}$, 
Fig. \ref{amp14} displays the exact BSW amplitude along with its asymptotic representation. 
In both cases the imaginary part
 dominates the real part and its zeros will produce oscillations in the differential 
cross section, as shown in Fig. \ref{sig14}. The
 differential cross section is given by
\begin{equation}
  \label{eq:sig1}
  \frac{d\sigma}{dt}=\frac{\pi}{s^2}|\mathcal{M}(s,\Delta)|^2\,,
\end{equation}
where for the asymptotic representation one uses Eq.~(\ref{eq:M4}). For both the BSW 
amplitude and its asymptotic representation, the real part has a local maximum near 
each zero of the imaginary part, and \textit{vice versa}. When the maximum of the real 
part near a zero of the imaginary part is relatively low, as in the case near 
$|t| = 0.5\text{GeV}^2$, one gets a sharp dip, but if not, like near 
$|t| = 2\text{GeV}^2$, one gets instead a smooth oscillation. Clearly the asymptotic 
result is larger than the exact BSW result, except near the diffraction peak, where 
they are hardly distinguishable. At a much higher energy $\sqrt{s} = 6000\text{TeV}$, 
the number of zeros increases, as shown in Fig. \ref{amp6000} and the low maximum of the real part
 near the zero of the imaginary part around $|t| = 0.2\text{GeV}^2$, generates a very
 sharp dip in the cross section, as seen in Fig. \ref{sig6000}, followed by another dip and some 
smooth oscillations.

\section{Concluding remarks}
\label{conclu}
\begin{figure}[htbp]
\begin{center}
\hspace*{-95mm}
  \begin{minipage}{6.5cm}
  \epsfig{figure=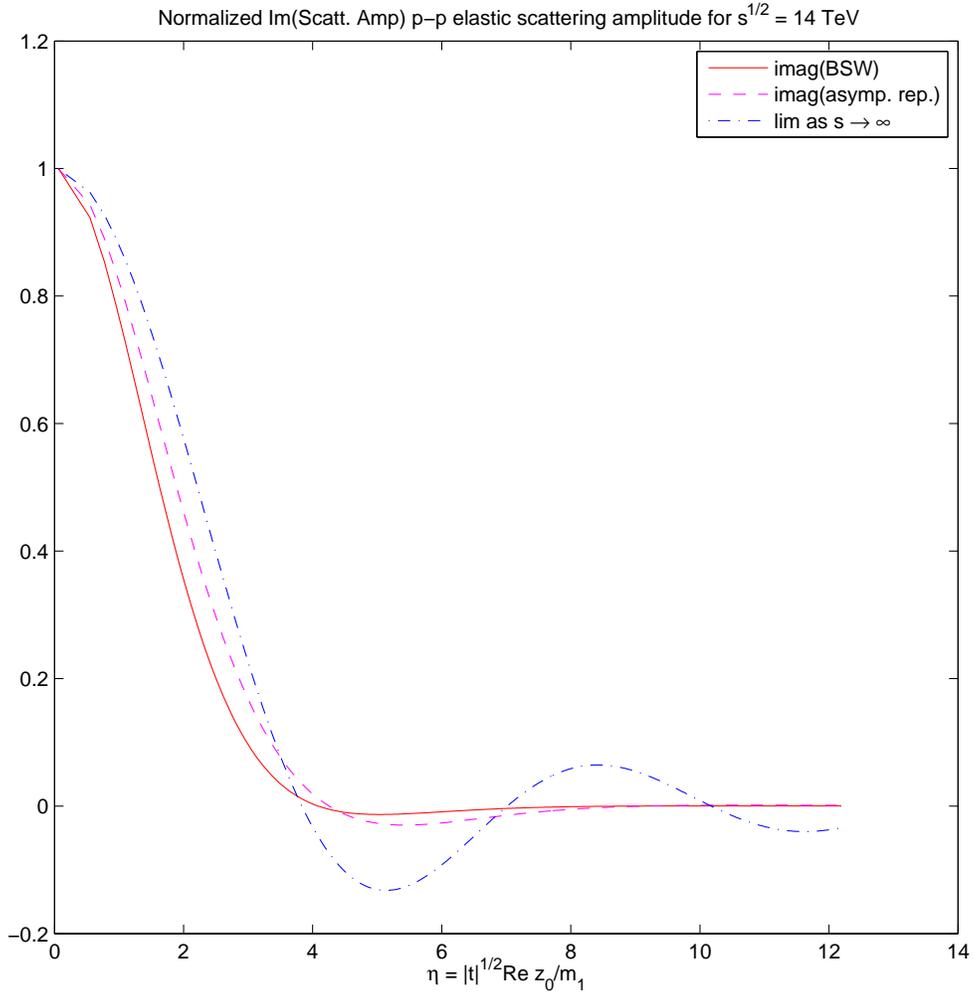,width=15.5cm}
  \end{minipage}
\end{center}
  \vspace*{-10mm}
\caption{
The normalized imaginary part of the elastic scattering amplitude for 
$\sqrt{s}= 14 \text{TeV}$. BSW solid, asymptotic dashed, lim$~s \rightarrow \infty$ dash-dotted.}
\label{sr14}
\vspace*{-1.5ex}
\end{figure}

\begin{figure}[htbp]
\begin{center}
\hspace*{-95mm}
  \begin{minipage}{6.5cm}
  \epsfig{figure=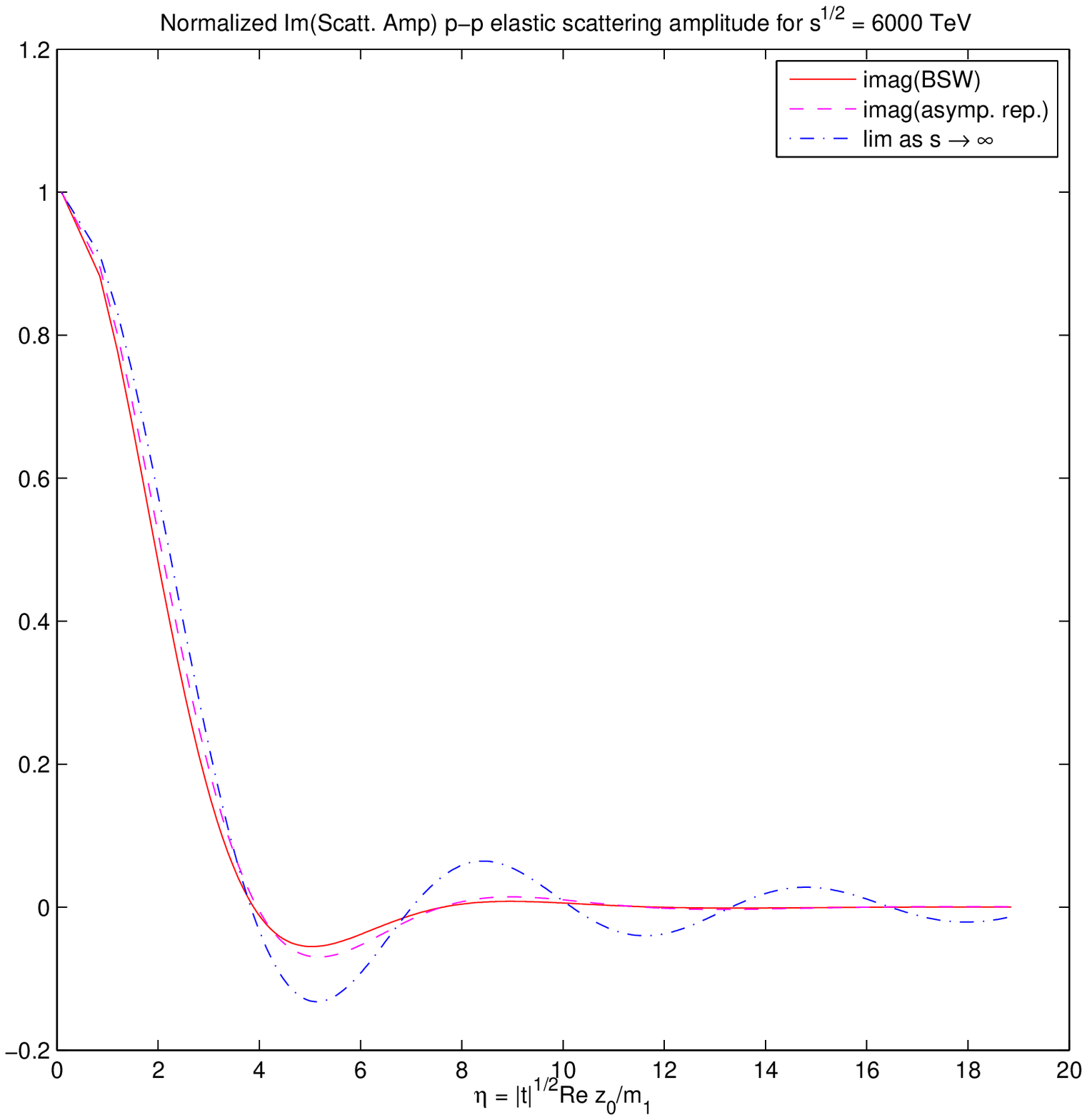,width=15.5cm}
  \end{minipage}
\end{center}
  \vspace*{-10mm}
\caption{
The normalized imaginary part of the elastic scattering amplitude for 
$\sqrt{s}= 6000 \text{TeV}$ (same legend as in Fig. \ref{sr14}).}
\label{sr6000}
\vspace*{-1.5ex}
\end{figure}

After recalling the basic features of the BSW model, we have presented our predictions for the LHC energies
and compared them with preliminary results from TOTEM. 
We have obtained the asymptotic representation of the BSW model in terms of a simple
formula. The existence of several zeros for both the real and the imaginary parts
of the scattering amplitude, generates oscillations in the differential cross section.
 The exact BSW result tends to coincide with this asymptotic representation, as
the energy increases. This is even more striking near the forward direction, 
in particular for the diffraction peak of the differential cross section. 
In connection with this, let us mention now
the following interesting feature of the asymptotic representation. A unitarity upper 
bound for the imaginary part of
the scattering amplitude for very high energy and small momentum transfers was 
derived a long time ago by Singh and Roy \cite{SR}. 
It reads \footnote{One should remember that the Singh-Roy amplitude is twice 
the BSW amplitude.}
\begin{equation}
  \label{eq:sr}
  \frac{\text{Im}\mathcal{M}(s,t)}{\text{Im}\mathcal{M}(s,0)} \leq\frac{2J_1(\sqrt{r})}{\sqrt{r}},~~~~~~~~\text{if}~~ r < 3.46~,
\end{equation}
with $r = |t|\sigma_{tot}(s)/4\pi$. For $s$ very large, from Eq.~(\ref{eq:M4}), 
we see that this variable is simply $\eta = \alpha \text{Re}z_0$. The ratio 
$\text{Im}\mathcal{M}(s,t)\over\text{Im}\mathcal{M}(s,0)$ has been plotted in Fig. \ref{sr14}, 
versus $\eta$, for $\sqrt{s} = 14\text{TeV}$. We compare the exact BSW result with 
the asymptotic representation and also with the Singh-Roy upper bound limit 
$s \to \infty$. We observe that the validity of the bound is, indeed, limited to 
$\eta < 3.46$. Fig. \ref{sr6000} displays the situation at 
$\sqrt{s} = 6000\text{TeV}$ and in this case the upper bound, whose validity is still 
limited to the diffraction peak, becomes much closer to the other two curves.\\

{\bf Acknowledgments} We thank J.M Richard for drawing our attention to the old bound 
of Singh and Roy. One of us (T.T.W) is greatly
indebted to the CERN Theory Group for their hospitality.

\appendix
\section{Appendix}
\setcounter{equation}{0}
\numberwithin{equation}{section}

For the small $\alpha$ region, we define 
\begin{eqnarray}
  \label{eq:Idef}
  \mathcal{I}_n(\beta)& = &\int_{-\infty}^\infty dx\,x^n\,\frac{d\;}{dx}
  e^{\displaystyle{-\beta e^{-x}}} \\&=&
\beta\int_{-\infty}^\infty dx\,x^n\,e^{-x}e^{\displaystyle{-\beta e^{-x}}}\\ &=&
\int_0^\infty dt [\ln(\beta/t)]^{n}e^{-t} \\ &=&
\left[\left(\ln \beta -\frac{d}{d\nu}\right)^n\int_0^\infty dt\,t^\nu e^{-t}\right]\bigg|_{\nu=0} \\ &=&
\left[\left(\ln \beta -\frac{d}{d\nu}\right)^n\Gamma(1+\nu)\right]\bigg|_{\nu=0},
\end{eqnarray}
where $t=\beta e^{-x}$ and $\Gamma$ is the Gamma function.
The results for $n= 0, 1, 2, 3$ are given in Table 2.
  Here $\gamma \approx 0.5772$ is the Euler's constant and $\zeta$ is the Riemann
  Zeta function.  The choice of $z_0$ made in Eq.~(\ref{eq:z0Def}) draws on the
  fact that
\begin{equation}
  \label{eq:I1}
\mathcal{I}_1(e^{-\gamma})= -\gamma +\gamma = 0, 
\end{equation}
hence Eq.~(\ref{eq:ch1}) follows as
\begin{eqnarray}
  \label{eq:ch}
  0&=&\int_{-\infty}^0 dx\,x\,\frac{d\;}{dx}e^{\displaystyle{-e^{-\gamma} e^{-x}}}
-\int_0^\infty dx\,x\,\frac{d\;}{dx}\left[1-e^{\displaystyle{-e^{-\gamma} e^{-x}}}\right] \\
&=& \int_0^\infty dx 
\left[1-e^{\displaystyle{-e^{-\gamma} e^{-x}}}\right] - \int_{-\infty}^0 dx\,e^{\displaystyle{-e^{-\gamma} e^{-x}}}~.
\end{eqnarray}

For the large $\alpha$ region one needs 
$\mathcal{J}(i\alpha)$ and $\mathcal{J}(-i\alpha)$ where we define
\begin{eqnarray}
  \label{eq:Jdef}
  \mathcal{J}(i\alpha)&\ = & \int_0^\infty dx\,e^{i\alpha x}
\left[1-e^{\displaystyle{-e^{-(x+\gamma)}}}\right] - \int_{-\infty}^0 dx\,e^{i\alpha
  x}e^{\displaystyle{-e^{-(x+\gamma)}}}\\ &=&
\frac{1}{i\alpha}\left[ -1+\int_{-\infty}^\infty e^{i\alpha x}\frac{d\;}{dx}
e^{\displaystyle{e^{-(x+\gamma}}}\right] \\ &=&
\frac{1}{i\alpha}\left[ -1+ e^{-i\gamma\alpha}\Gamma(1-i\alpha)\right]~. \label{eq:A11}
\end{eqnarray}
From Eq.~(\ref{eq:A11}) and the properties of the Gamma function \cite{bI}, it
follows that $\mathcal{J}(-i\alpha)$ is the complex conjugate of
$\mathcal{J}(i\alpha)$.\\ For small values of $\alpha$ one finds
\begin{equation}
  \label{eq:Jsmall}
  \mathcal{J}(i\alpha)\approx \frac{i\pi^2\alpha}{12}-\frac{\alpha^2}{3}\zeta(3)~.
\end{equation}
 \begin{table}
    \centering
    \begin{tabular}{|c|l|}\hline
$n$ & $\mathcal{I}_n(\beta)$\\ \hline
$0$& $1$\\
$1$&$\ln\beta+\gamma$\\
$2$&$(\ln\beta+\gamma)^2+\pi^2/6$\\
$3$&$(\ln\beta+\gamma)^3+\frac{\pi^2}{2}(\ln\beta+\gamma)+2\zeta(3)$\\ \hline
    \end{tabular}
    \caption{\label{tab:table2} Values of $\mathcal{I}_n(\beta)$}
  \end{table}

\clearpage
\newpage
\newpage

\newpage

\newpage

\newpage

\newpage
\clearpage
\newpage

\newpage

\clearpage
\newpage

\newpage

\newpage


\newpage
\begin{thebibliography}{99}

\bibitem{chengWu}  H. Cheng and T. T. Wu, Phys. Rev. Lett. \textbf{24}, 1456
  (1970) (See also H. Cheng and T. T. Wu, \textit{Expanding Protons: Scattering at
    High Energies} (MIT Press, Cambridge, MA, 1987)).
\bibitem{BSW79} C. Bourrely, J. Soffer, and T. T. Wu, Phys. Rev. D \textbf{19}, 3249
  (1979).
\bibitem{BSW84} C. Bourrely, J. Soffer, and T. T. Wu, Nucl. Phys. B
  \textbf{247}, 15 (1984).
 \bibitem{BSW03} C. Bourrely, J. Soffer, and T. T. Wu, Eur. Phys. J. C
  \textbf{28}, 97 (2003).
\bibitem{block11} M. M. Block and F. Halzen, Phys. Rev. D \textbf{83} (2011) 077901.
\bibitem{islam09} M. M. Islam et \textit{al}., Int. J. Mod. Phys. A \textbf{21} 
(2006) 1.
\bibitem{kaspar11} J. Ka\v{s}par et \textit{al}., Nucl. Phys. B \textbf{843} (2011)
 84.
\bibitem{carvalho05} P.A.S. Carvalho et  \textit{al}., Eur. Phys. J. C \textbf{39} 
(2005) 359.
\bibitem{petrov03} V. Petrov et \textit{al}., Eur. Phys. J. C \textbf{28} (2003) 
525.
\bibitem{totem} G. Antchev et \textit{al}., Eur. Phys. Lett.  \textbf{95} (2011) 
41001, \textbf{96} (2011) 21002.
\bibitem{alfa} ATLAS-ALFA collaboration, CERN-LHCC-2004-010,  CERN-LHCC-2008-004.
\bibitem{BSW11} C. Bourrely, J. Soffer, and T. T. Wu, Eur. Phys. J. C
  \textbf{71}, 1061 (2011).
  \bibitem{bII}A. Erd\'elyi, ed., \textit{Higher Transcendental Functions}, Vol
  II (McGraw-Hill Book Co., New York, 1953).
  \bibitem{SR}  V. Singh and S.M. Roy, Phys. Rev. D \textbf{1}, 2638
  (1970).
\bibitem{bI}A. Erd\'elyi, ed., \textit{Higher Transcendental Functions}, Vol
  I (McGraw-Hill Book Co., New York, 1953).
\end{thebibliography}
\end{document}